\documentclass[a4paper,onecolumn]{article}
\usepackage{amsmath}
\usepackage{graphicx}
\usepackage{braket}
\usepackage{caption}
\usepackage{subcaption}
\usepackage{authblk}
\usepackage{verbatim}

\title{String picture of 1+1 dimensional QED in light--front formulation
\thanks{Presented at Light Cone 2012, 8-13 July 2012, Polish Academy of Arts and Sciences, Cracow}}
\author[1,2]{Zbigniew Ambrozi\'nski\thanks{zbigniew.ambrozinski@uj.edu.pl}}
\affil[1]{Marian Smoluchowski Institute of Physics, Jagiellonian University\\Reymonta 4, 30-059 Cracow, Poland}
\affil[2]{Max-Planck-Institut f\"ur Gravitationsphysik (Albert-Einstein-Institut)\\Am M\"uhlenberg 1, D-14476 Golm, Germany}

\begin{document}
\maketitle

\begin{abstract}
We study quantum electrodynamics in 1+1 dimensions in the light--front frame using numerical methods. We analyze confinement and charge screening which are key features of this system. By direct analysis of wavefunctions of bound states in two parton sector we determine the string tension. In four parton sector we introduce inclusive distributions and inspect structure of energy eigenstates. We conclude that they are composed of two weakly interacting $f\bar f$ pairs. These four particle states are responsible for the screening. Finally, we study time evolution of a fermion--antifermion state separated by a specific distance. We demonstrate that for sufficient separation it decays into a multiparton state and the number of particles in the product depends on separation of particles.
\end{abstract}

\section{Introduction}
Quantum electrodynamics in 1+1 dimensions is the simplest nontrivial gauge field theory. Still, it retains some important features of 4--dimensional QCD, confinement being the most remarkable. For these reasons it has been extensively studied in many aspects. The model is soluble in two limits. For zero mass it becomes a theory of free bosons \cite{Schwinger} while for vanishing coupling it is a free Dirac theory. Both approximations, small and large mass were studied eg. by Coleman \cite{Coleman}. An approach using perturbation theory in the small mass limit was presented in \cite{Adam}.

Apart from analytical methods, the Schwinger model was also solved on the lattice, eg. by Schiller \cite{Schiller}. Another successful technique the light--front frame quantization. It was first proposed by Susskind \cite{Susskind} and developed by Chang et al. \cite{Chang}. First application of the light--front formalizm to the two dimensional Schwinger model was performed by Brodsky et al. in \cite{Pauli,Eller}. This formulation opened the possibility to compute energies of the system by numerical methods for any mass parameter $m$. As the method proved useful in this simple case it was later applied to more complex theories as QCD in 3+1 dimensions.

A well known feature of the massive Schwinger model is that for small coupling constant the system may be well approximated by a fermion and antifermion with a linear potential between them \cite{Coleman}. Linearity of the potential heavily relies on having only one spatial dimension. Due to this fact fermions are confined. The string tension, which is the proportionality constant between energy end separation of partons, was found eg. in \cite{Gross,Hamer,Kazakov}. It is also known \cite{Gross} that the interaction of largely separated fermion--antifermion pair is screened due to vacuum polarization.

The aim of this paper is to show how above results can be read directly from numerical data obtained in the light--front quantization. In particular, we get the linear energy dependence and determine the string tension. We also study the structure of 4--particle bound states which are responsible for the screening. In order to handle the problem of visualizing probability distribution of 4--particle state we use inclusive which are widely used rather in the context of scattering. Finally, we demonstrate how two charges at large distances are screened by vacuum polarization.

Outline of the paper is as follows. In sect. 2 we present the light--front quantization method of the massive Schwinger model. In sect. 3 we show numerical results for masses and construction of wavefunctions. Sect. 4 is devoted to considerations in the 2--particle sector where the linear potential emerge. In sect. 5 we discuss the structure of bound states with 4--particle components responsible for the screening. In sect. 6 we demonstrate that a widely separated $f\bar f$ pair indeed decays into multi particle state. Results are summarized in sect. 7.

\section{The model}
The theory is quantized following closely \cite{Eller} and their notation. The Lagrangian of the massive Schwinger model reads
\begin{align}
\mathcal L=i\bar\psi\gamma^\mu\partial_\mu\psi-m\bar\psi\psi-\frac{1}{4}F^{\mu\nu}F_{\mu\nu}-g\bar\psi\gamma^\mu\psi A_\mu.
\end{align}
We work in the light--front coordinates $x^\pm=x^0\pm x^1$ and in the light--front gauge $A^+=0$. The only independent field is $\psi_+(x)\equiv \Lambda^{(+)}\psi(x)$ where $\Lambda^{(+)}=\frac{1}{4}\gamma^-\gamma^+$ is a projection operator. The field is quantized at the equal light--front time $x^+$:
\begin{align}
\{\psi_+(x),\psi_+(y)\}_{x^+=y^+}=\Lambda^{(+)}\delta(x^--y^-)
\end{align}
with periodic boundary conditions. Field $\psi_+(x)$ is expanded in Fourier modes:
\begin{align}
\psi_+(x)=\frac{u}{\sqrt{2L}}\sum_{k=1}^{\infty}(b_ke^{-i\pi kx^-/L}+d^\dagger_ke^{i\pi kx^-/L}).
\end{align}
Operators $b_k^\dagger$ and $d_k^\dagger$ create fermion $f$ and antifermion $\bar f$ with light--front momentum $p^+=2\pi k/L$ respectively. $u$ is a spinor satisfying $\Lambda^{(+)}u=u,\ u^\dagger u=1$.
There are three conserved quantities, the charge $Q$, the light--front momentum $P^+$ and the hamiltonian $P^-$:
\begin{align}
Q&=\sum_{k=1}(b^\dagger_kb_k-d^\dagger_kd_k),\\
P^+&=\frac{2\pi}{L}K\equiv\frac{2\pi}{L}\sum_{k}k(b^\dagger_kb_k+d^\dagger_kd_k),\\
P^-&=\frac{L}{2\pi}(m^2H_0+\frac{g^2}{\pi}V),
\end{align}
where the specific form of a free hamiltonian $H_0$ and potential $V$ is given in \cite{Eller}. Charged states have infinite energy for any values of $g$, so we consider only states with zero charge, hence equal number of fermions and antifermions. In the continuum limit the light--front resolution $K$ and size of the space $L$ are infinite while $P^+$ is kept finite. Our aim is to find eigenvalues and eigenstates of the invariant mass operator $M^2=P^+P^-$.

\section{Numerical results}
\begin{figure}[htb]
\centering
\includegraphics[width=.8\textwidth]{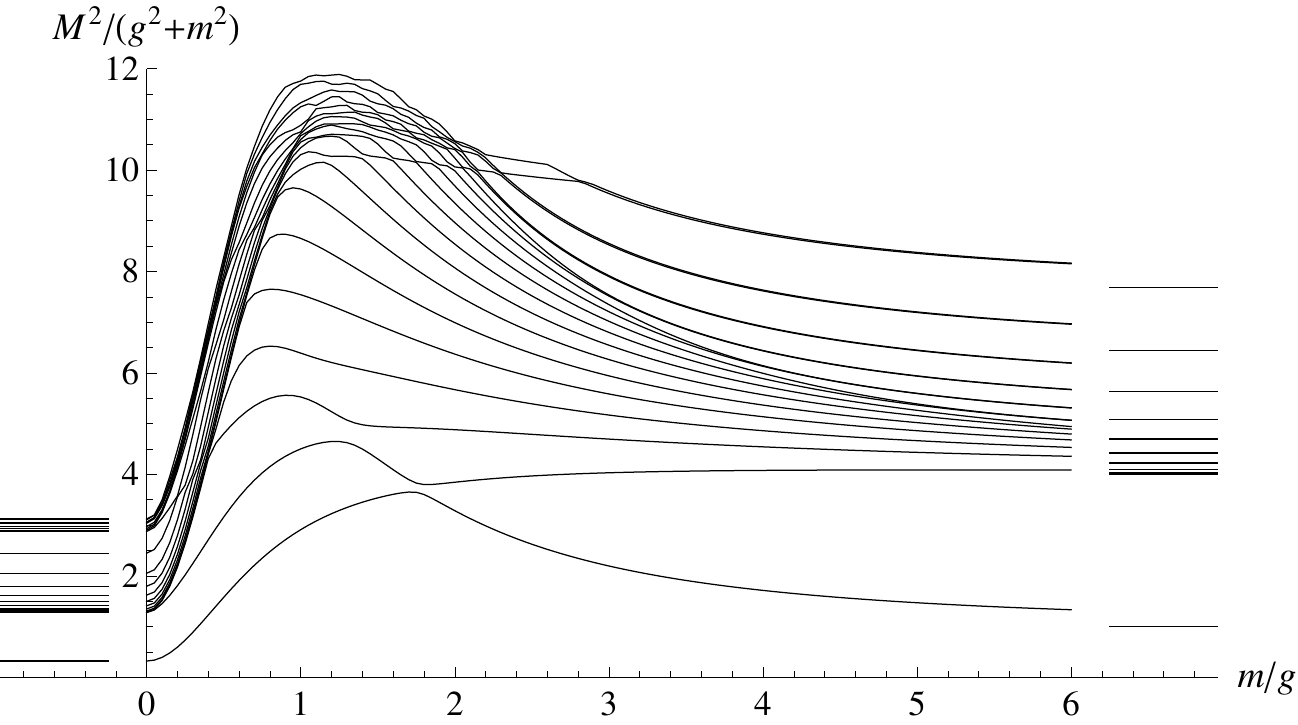}
\caption{First 20 eigenvalues of the $M^2$ matrix as functions of $m/g$. The light-front resolution is $K=25$. Lines on the left and right of the plot are masses in the massless and free limit respectively.}\label{fig:energies}
\end{figure}

We diagonalize $M^2$ on eigenspaces of fixed light--front resolution $K$ which is possible since $M^2$ and $K$ commute. Basis of the Fock space consists of states of the form $\ket{\{k_i\},\{\bar k_j\}}=\ket{k_1,\ldots,k_N,\bar k_1,\ldots,\bar k_N}\equiv b^\dagger_{k_1},\ldots,b^\dagger_{k_N},d^\dagger_{\bar k_1},\ldots,d^\dagger_{\bar k_N}\ket{0}$. Due to the relation $K=\sum k_i$, $K$ simultaneously limits parton number $k_i$ and number of partons $2N$, hence size of the Hilbert space. Therefore, $M^2$ becomes a finite matrix and can be diagonalized numerically. Physical values of mass are obtained in the limit $K\to\infty$. The matrix $M^2$ does not depend on $L$. In Fig. \ref{fig:energies} we present the lower spectrum of $M^2$ for $K=25$ as a function of a fermion mass $m$. The spectrum coincides with spectra of soluble models in limits $m\to 0$ and $m\to \infty$. Plotted eigenvalues already converged with $K$.

Given the matrix of the mass operator, we obtain not only eigenvalues but also eigenvectors. From these we can read wavefunctions of bound states. A state $\ket{\phi}$ can be written as a superposition of states with definite number of particles
\begin{align}
\ket{\phi}&=\sum_{N=1}^{N_{max}}\ket{\phi_{2N}}\\
\ket{\phi_{2N}}&=\sum_{k_i,\bar k_j}\alpha_{2N}(k_i,\bar k_j)\ket{k_i,\bar k_j},&i,j=1,\ldots,N
\end{align}
Note that momenta do not have proper dimensions.
Coefficients $\alpha$ can be read from eigenvectors of the matrix. Wavefunction $\phi_{2N}(k_i,\bar k_j)$ of state $\ket{\phi_{2N}}$ is given by coefficients $\alpha_{2N}$ antisymmetrized under the transformations $k_i\leftrightarrow k_j$ and $\bar k_i\leftrightarrow \bar k_j$. Notice that the momentum of the $i$--th parton $k_i$ is dimensionless and corresponds to physical momentum $p^+_i=2\pi k_i/L$. A wavefunction in momentum space is give by the Fourier transform:
\begin{align}
\phi_{2N}(x^-_i,\bar x^-_j)=\sum_{k_i,\bar k_j}\exp\left(-\frac{\pi i}{L}\sum_n(x^-_n k_n+\bar x^-_n\bar k_n)\right)\phi_{2N}(k_i,\bar k_j).
\end{align}
Since $K$ is fixed and $\sum_i(k_i+\bar k_i)=K$, the wavefunction depends only on differences of positions, while dependence on one coordinate is trivial. This coordinate can be chosen arbitrarily and we choose $\bar x_N$. Then,
\begin{align}
\phi_{2N}(x^-_i,\bar x^-_j)=e^{-\frac{i}{2}P^+\bar x^-_N}\phi_{2N}(\Delta_i,\bar \Delta_j),
\end{align}
where $\Delta_i=x^-_i-\bar x^-_N,\ i=1,\ldots,N$ and $\bar \Delta_j=\bar x^-_j-\bar x^-_N,\ j=1,\ldots,N-1$.

In what follows we will be interested only in wavefunctions in 2 and 4 parton sectors, i.e. $N=1,2$.

\section{Two particle sector}
\begin{figure}[htb]
\centering
\includegraphics[width=\textwidth]{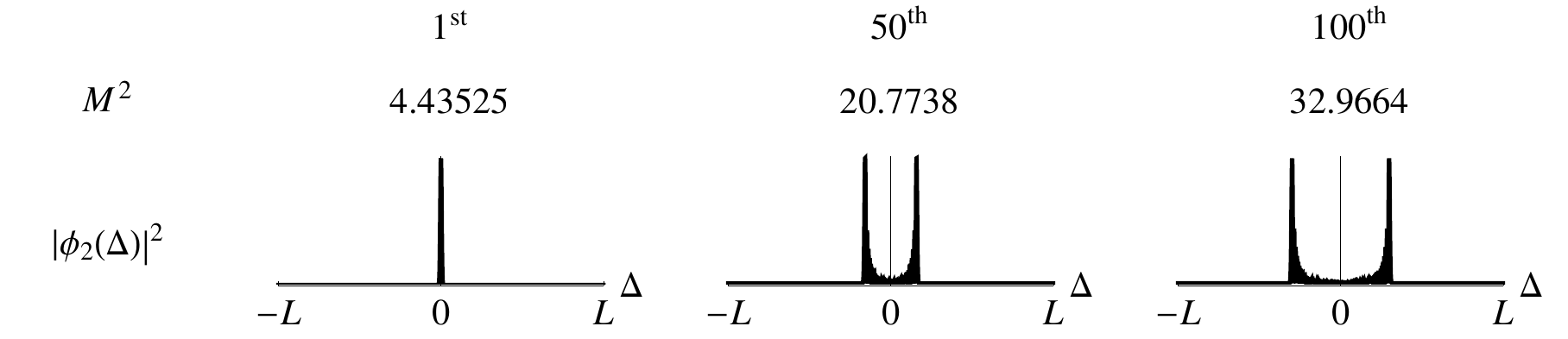}
\caption{Probability distribution of $\Delta$ for different energy states. For high states width of peaks at $\pm\Delta^*$ is small compared to $\Delta^*$. Plot was made for $m=1,\ g=0.3$. For smaller $g$ peaks are wider and the function $|\phi_2(\Delta)|^2$ is nonzero inside the interval $(-\Delta^*,\Delta^*)$.
}\label{fig:two_particle_sector}
\end{figure}

The wavefunction of two partons can be written as
\begin{align}
\phi_2(x^-,\bar x^-)=e^{-\frac{i}{2}P^+\bar x^-}\phi_2(\Delta),
\end{align}
where $x^-$ is position of a single fermion $f$,  $\bar x^-$ is position of an antifermion $\bar f$ and $\Delta=x^--\bar x^-$. Plot of $|\phi_2(\Delta)|^2$ for several states is presented in Fig. \ref{fig:two_particle_sector}. The probability distribution $|\phi_2(\Delta)|^2$ has two sharp peaks at $\Delta=\pm\Delta^*$, whose width is proportional to $1/P^+$. For highly excited states widths of peaks are small compared to $\Delta^*$ and therefore the distance between two particles is well defined. Thus $M^2$ can be represented as a function of $\Delta^*$. It turns out that the relation is linear. A linear fit yields $M^2\approx m_0^2+\frac{1}{2}P^+g^2\Delta^*$. It reflects linearity of the potential $V$ in the two particle sector.

\section{Four particle sector}
Let us now con consider 4--particle sector. The wavefunction again can be written as
\begin{align}
\phi_4(x_1^-,x_2^-,\bar x_1^-,\bar x_2^-)=e^{-\frac{i}{2}P^+\bar x_2^-}\phi_4(\Delta_1,\Delta_2,\bar \Delta_1).
\end{align}
Variables $x_1^-, x_2^-$ are positions of fermions and $\bar x_1^-,\bar x_2^-$ are positions of antifermions. We construct inclusive distributions \cite{Dorigoni} of the form
\begin{align}
D(\Delta)=\int d\Delta_1d\Delta_2d\Delta_3\delta(\Delta-|\Delta_{ij}|)|\phi_4(\Delta_1,\Delta_2,\Delta_3)|^2,
\end{align}
where $\Delta_{ij}=x_i^--x_j^-$ is the relevant distance. We are particularly interested in the following profiles:
\begin{itemize}
\item $D_{ff}(\Delta)$, where $\Delta_{ij}=\Delta_{12}=x^-_1-x^-_2$ is the relative distance between two fermions $f$,
\item $D_{f\bar f}(\Delta)$, where $\Delta_{ij}=\Delta_{1\bar1}=x^-_1-\bar x^-_1$ is the distance between a fermion $f$ and an antifermion $\bar f$; due to antisymmetry of $\psi_4(x_1,x_2,\bar x_1,\bar x_2)$ particular choice of fermion and antifermion is arbitrary,
\item $D_{f\bar f}^{nn}(\Delta)$, where $\Delta_{ij}=\Delta_{1\bar 1}$ and an additional factor $\Theta(|\Delta_{1\bar 2}|-|\Delta_{1\bar 1}|)$ under the integral is present; this profile is the probability distribution of the distance between a fermion and the nearest antifermion.
\end{itemize}

\begin{figure}[htb]
\centering
\includegraphics[width=.9\textwidth]{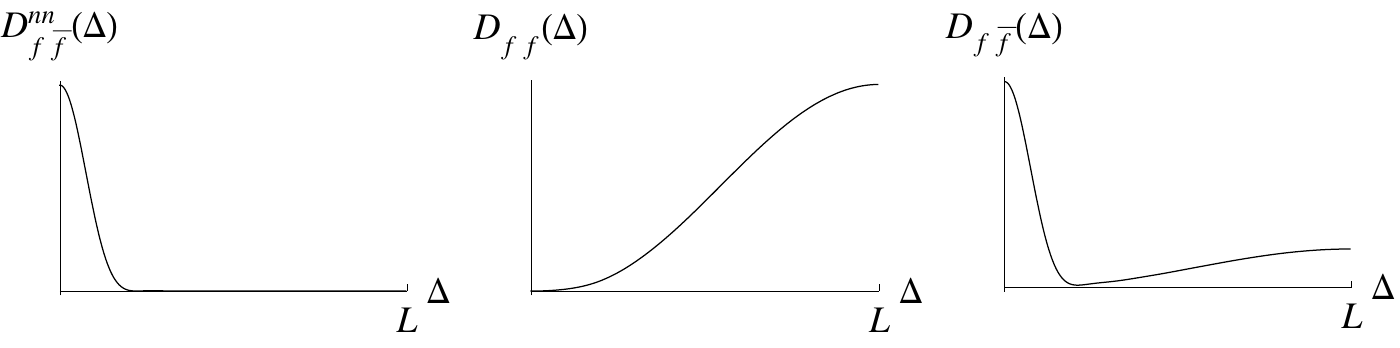}
\caption{Four particle inclusive distributions for 7--th state of the spectrum for $m/g=1$. This is the lowest state with non--negligible contribution from four particle sector. The contribution from two particle sector is smaller than 0.1\%.}\label{fig:plots7}
\includegraphics[width=.9\textwidth]{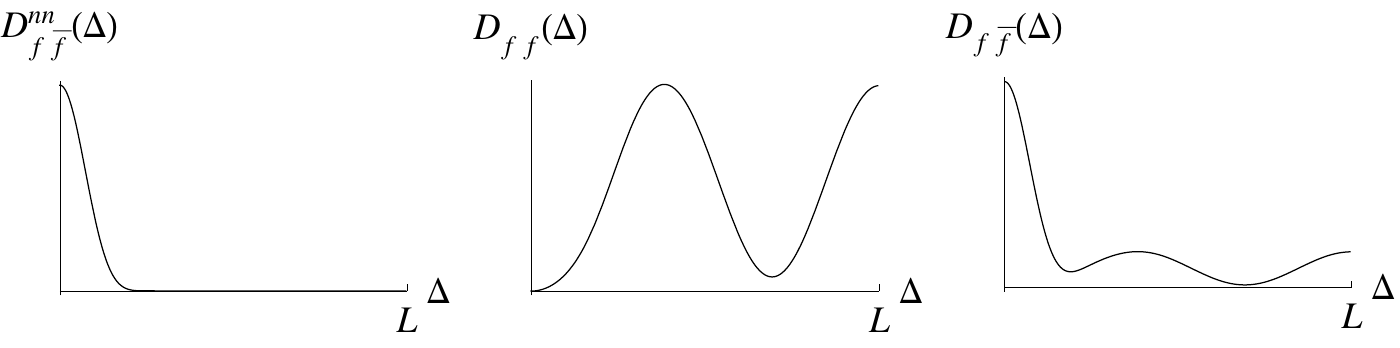}
\caption{Four particle inclusive distributions for 8--th state of the spectrum for $m/h=1$. The contribution from two particle sector is smaller than 1\%.}\label{fig:plots8}
\end{figure}

A priori the eigenvectors of matrix $M^2$ can have components with arbitrary parton numbers. However, we observe that several lowest mass states are almost exclusively composed of the two particle states. Next, there is a sequence of states in four particle sector. For higher $g$ the binding energy at given distance is larger. Therefore, smaller distance is needed to reach energy required to create an additional pair and the four particle sector appears earlier.

Inclusive distributions for first and second eigenstates in four particle sectors are shown in Figs. \ref{fig:plots7} and \ref{fig:plots8}. In both cases the function $D_{f\bar f}^{nn}(\Delta)$ is highly peaked at the origin and vanishes almost exactly elsewhere. It means that the fermion $f$ at $\Delta=0$ forms a pair with an antifermion $\bar f$. Function $D_{ff}(\Delta)$ is zero at the origin. This is reflection of the Pauli exclusion principle. For the lowest energy state in four particle sector $D_{ff}(\Delta)$ weakly depends on $\Delta$. For higher states it grows faster for small $\Delta$ and exhibits oscillatory behavior for larger $\Delta$. Function $D_{f\bar f}(\Delta)$ is peaked at $\Delta=0$. This peak is directly related to the peak of $D_{f\bar f}^{nn}(\Delta)$. Then, it has maxima approximately at the same positions at which $D_{ff}(\Delta)$ has maxima. These correspond to an antifermion $\bar f$ which forms a pair with the other fermion $f$.
The dependence of $D_{ff}(\Delta)$ on $\Delta$ is so weak because the two $f\bar f$ pairs are neutral and interact indirectly only due to the exclusion principle among constituents. Therefore, they can move almost independently.

\section{Decay of $f\bar f$ state.}
\begin{figure}[htb]
\centering
\begin{subfigure}{.48\textwidth}
\centering
\includegraphics[width=\textwidth]{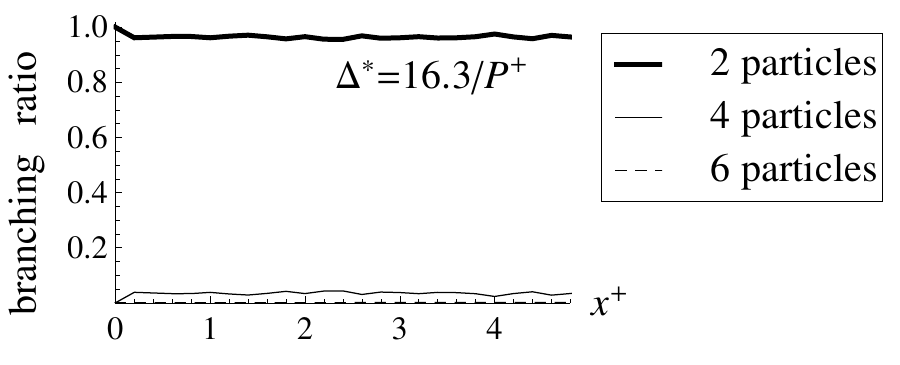}
\end{subfigure}\quad
\begin{subfigure}{.48\textwidth}
\centering
\includegraphics[width=\textwidth]{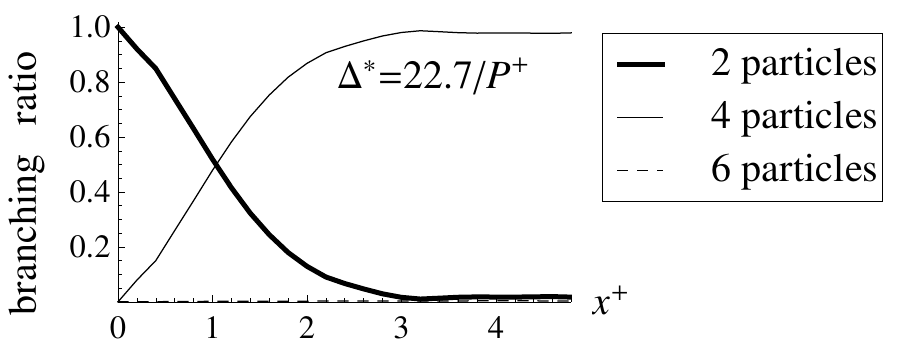}
\end{subfigure}\\
\begin{subfigure}{.5\textwidth}
\centering
\includegraphics[width=\textwidth]{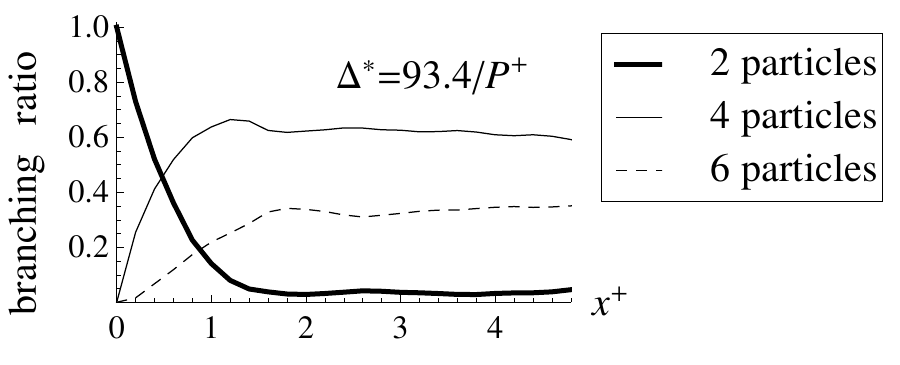}
\end{subfigure}
\caption{Evolution of distribution of number of particles. Initial state consists of $f\bar f$ pair at given distance $\Delta^*$. Values of $x^+$ are given in units $P^+/g^2$. Plots are made for $m/g=0.5$. For small distance the state remains in the two particle sector. For larger $\Delta^*$ it decays into four particles. For $\Delta^*=93.4/P^+$ it decays into four particles.
}\label{fig:evolutions}
\end{figure}

From above considerations we infer that when a $f\bar f$ pair is separated by a distance which is large enough, there exist a four particle state composed of two $f\bar f$ pairs which has smaller energy. Then a single pair can decay into two pairs. We checked it explicitly using evolution in the light--front time.
Let us construct a pair of $f$ and $\bar f$ separated by a distance $\Delta^*$. This is obtained by taking a mass eigenstate in two particle sector, for which the separation is well defined. Then let us evolve it in the light--front time $x^+$ using the hamiltonian $P^-$. Finally, we plot contribution of each multi--particle sector (branching ratio) to the evolved state as a function of $x^+$. Results are presented in Fig. \ref{fig:evolutions}.

One can see that for small distance $\Delta^*$ the state remains in the two--particle sector. This is because each state with at least 4 partons has higher energy than the initial state. The potential energy of the two particles is not high enough to create additional fermions. For $\Delta^*=22.7/P^+$ the initial $f\bar f$ pair decays into four particle state. The half--life time can be read from the plot and is $x^+=1.04 P^+/g^2$ for the ratio $m/g=0.5$. For yet larger distance the initial state decays into four and six particles. The half--time of the $f\bar f$ particle is much shorter since the potential energy is larger.

We conclude that a $f\bar f$ pair can decay into multi--particle states if the separation of partons is large enough and subsequently potential energy is sufficient. The number of particles into which it decays as well as the halt--life depends on separation $\Delta^*$.

\section{Conclusions}
Summarizing, the inclusive distributions give us a better insight into the structure of bound states of the multiparton system. We observed that in the two--particle sector the distance between fermion and antifermion is well defined for bound states. We established relation between invariant mass and separation, showed that it is linear and extracted the string tension. Then we investigated low energy bound states with four particles. It turned out that they consist of two $f\bar f$ pairs which are almost independent. Finally, we conjectured that a pair can decay into multi--particle state whenever its potential energy is sufficient. This statement was confirmed by direct evolution of single pairs with different separations $\Delta^*$. If the separation is large enough more pairs are created. The time of creation and number of created particles depends on the initial potential energy. Additional particles screen the interaction of the initial pair.

\section{Acknowledgements}
This work was supported by the Foundation for Polish Science MPD Programme co-financed by the EU European Regional Development Fund, agreement no. MPD/2009/6.

\begin{thebibliography}{10}
\bibitem{Schwinger}
  J.~S.~Schwinger,
  Phys.\ Rev.\  {\bf 128}, 2425 (1962).

\bibitem{Coleman}
  S.~R.~Coleman,
  Annals Phys.\  {\bf 101}, 239 (1976).

\bibitem{Adam}
  C.~Adam,
  Annals Phys.\  {\bf 259}, 1 (1997)
  [hep-th/9704064].

\bibitem{Schiller}
  A.~Schiller and J.~Ranft,
  Nucl.\ Phys.\ B {\bf 225}, 204 (1983).

\bibitem{Susskind}
  L.~Susskind,
  Phys.\ Rev.\  {\bf 165}, 1535 (1968).

\bibitem{Chang}
  S.~-J.~Chang, R.~G.~Root and T.~-M.~Yan,
  Phys.\ Rev.\ D {\bf 7}, 1133 (1973).

\bibitem{Pauli}
  H.~C.~Pauli and S.~J.~Brodsky,
  Phys.\ Rev.\ D {\bf 32}, 1993 (1985).

\bibitem{Eller}
  T.~Eller, H.~C.~Pauli and S.~J.~Brodsky,
  Phys.\ Rev.\ D {\bf 35}, 1493 (1987).

\bibitem{Gross}
  D.~J.~Gross, I.~R.~Klebanov, A.~V.~Matytsin and A.~V.~Smilga,
  Nucl.\ Phys.\ B {\bf 461}, 109 (1996)
  [hep-th/9511104].

\bibitem{Hamer}
  C.~J.~Hamer, J.~B.~Kogut, D.~P.~Crewther and M.~M.~Mazzolini,
  Nucl.\ Phys.\ B {\bf 208}, 413 (1982).

\bibitem{Kazakov}
  V.~A.~Kazakov and I.~K.~Kostov,
  Nucl.\ Phys.\ B {\bf 176}, 199 (1980).

\bibitem{Dorigoni}
  D.~Dorigoni, G.~Veneziano and J.~Wosiek,
  JHEP {\bf 1106}, 051 (2011)
  [arXiv:1011.1200 [hep-th]].



\end{thebibliography}

\end{document}